\documentclass{article}
\usepackage{ijcai25}

\usepackage{times}
\usepackage{soul}
\usepackage{url}
\usepackage[hidelinks]{hyperref}
\usepackage[utf8]{inputenc}
\usepackage[small]{caption}
\usepackage{graphicx}
\usepackage{amsmath}
\usepackage{amsthm}
\usepackage{booktabs}
\usepackage{algorithm}
\usepackage{algorithmic}
\usepackage{amsfonts}
\usepackage[switch]{lineno}
\usepackage{comment}

\title{M3G: Multi-Granular Gesture Generator for Audio-Driven Full-Body Human Motion Synthesis}

\author{
Zhizhuo Yin$^1$
\and
Yuk Hang Tsui$^2$\and
Pan Hui$^{1,2}$\\
\affiliations
$^1$The Hong Kong University of Science and Technology (Guangzhou)\\
$^2$The Hong Kong University of Science and Technology\\
\emails
zyin190@connect.hkust-gz.edu.cn,\\
yhtsui@connect.ust.hk,\\
panhui@ust.hk
}

\begin{document}

\maketitle

\begin{abstract}
Generating full-body human gestures encompassing face, body, hands, and global movements from audio is a valuable yet challenging task in virtual avatar creation. Previous systems focused on tokenizing the human gestures framewisely and predicting the tokens of each frame from the input audio. However, one observation is that the number of frames required for a complete expressive human gesture, defined as granularity, varies among different human gesture patterns. Existing systems fail to model these gesture patterns due to the fixed granularity of their gesture tokens. To solve this problem, we propose a novel framework named Multi-Granular Gesture Generator (M3G) for audio-driven holistic gesture generation. In M3G, we propose a novel Multi-Granular VQ-VAE (MGVQ-VAE) to tokenize motion patterns and reconstruct motion sequences from different temporal granularities. Subsequently, we proposed a multi-granular token predictor that extracts multi-granular information from audio and predicts the corresponding motion tokens. Then M3G reconstructs the human gestures from the predicted tokens using the MGVQ-VAE. Both objective and subjective experiments demonstrate that our proposed M3G framework outperforms the state-of-the-art methods in terms of generating natural and expressive full-body human gestures. 
\end{abstract}

\section{Introduction}

\begin{figure*}[!htb]
\centering
\includegraphics[width=0.8\textwidth]{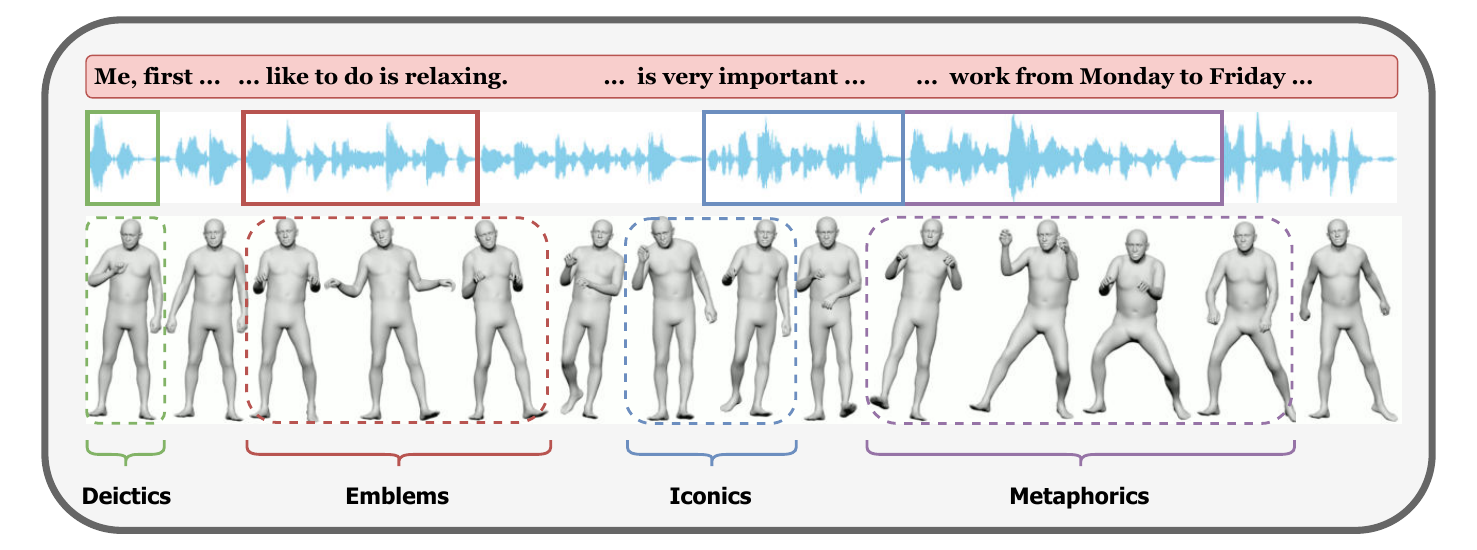} 
\caption{Illustration of gesture sequences with expressive motion patterns in different granularities. }
\label{granularity illustration}
\end{figure*}

Among all human communication, the movements of all body gestures serve as an important approach to conveying thoughts. Such non-verbal signals provide more information other than voice and context, enhancing the expressiveness and vividity of the speech, thus allowing listeners to gain a more comprehensive understanding of the intentions, emotions, and motivations of the speaker~\cite{tipper2015body,pelachaud2009studies}. Existing studies~\cite{van1998persona,10.1145/2668064.2668102} suggest that expressive gestures make avatars more intimate and trustworthy. In the metaverse~\cite{lee2021all}, the naturalness of the avatar’s full-body gestures, such as face micro-expression, intricate body gestures, and movement trajectory, could impact the sense of realism, presence, and satisfaction. Therefore, generating natural full-body gestures is a valuable and challenging task that serves as a critical component for creating realistic digital humans~\cite{shiohara2023blendface,zhou2022audio}. 

Traditional methods~\cite{cassell2001beat,huang2012robot} utilized rule-based algorithms to blend different preset motion clips according to the speech. Such workflow suffers from inconsistency between the audio and motion, and the unnatural transition among different motion clips. Recent studies attempt to address this problem by leveraging deep learning models to map the audio~\cite{ahuja2020no,habibie2021learning} and text~\cite{ahuja2019language2pose,bhattacharya2021text2gestures} information to human pose sequences. However, the process of mapping the information into continuous latent space and generating corresponding motions is highly sensitive to noise signals, which poses challenges for practical implementation in real-world scenarios. Thus, the most recent work~\cite{liu2023emage,ao2022rhythmic,yazdian2022gesture2vec} focuses on utilizing VQ-VAE~\cite{van2017neural}, a method that encodes continuous motion latent representations into a discrete latent space while preserving the original motion information, in order to generate human motions. These methods first utilize VQ-VAE to learn the patterns of human poses in each frame and project them into discrete latent space, represented by motion tokens. Then, they utilize their token predictor network to transform the audio and text representations into the motion token series and reconstruct human motion by applying the decoder from VQ-VAE. 

However, in these methods, one shared assumption is that each motion token exclusively represents a static human pose in a single frame. Since natural gestures convey semantic information by a series of sequential poses with varying durations, shown in Figure~\ref{granularity illustration}, this assumption severely hampers the ability of VQ-VAE to learn natural gestures that express intricate semantic information. Such an assumption also constrains the capability of the token predictor model to capture expressive patterns of audio and transcript representations with inconsistent durations and generate accurate corresponding motion tokens.

To solve the above problem, we propose the concept of granularity in this audio-driven generation model, which is defined as the number of frames for a complete expressive gesture. We further propose the Multi-Granular Gesture (M3G) framework, which models the motion patterns in different granularities and reconstructs dynamic, holistic body gestures based on multi-granular motion representations.

One of the key components of M3G is the Multi-Granular VQ-VAE (MGVQ-VAE) model, which encodes the motion patterns with different granularities into a shared discrete latent space as a series of motion tokens and reconstructs the motion sequence from these motion tokens.
Experiments show that the motion reconstructed from a series of token series with multiple granularities performs better than existing single-granular VQ-VAE methods. This proves that MGVQ-VAE demonstrates superior ability in modeling diverse motion patterns, particularly those with varied durations, and enhances performance in reconstructing expressive and dynamic body motions. 

In audio-driven motion generation, both input audio and transcripts are transformed into latent space represented by embeddings and utilized to predict motion token series with different granularities. To cope with MGVQ-VAE, we proposed a Multi-Granular Token Predictor that can capture sequential patterns from audio and transcript and generate motion token series with multiple granularities.

To summarize, the main contributions are as follows:

\begin{itemize}
\item This work introduces a novel framework called the Multi-Granular Gesture Generator (M3G) which generates dynamic and natural full-body gestures using multi-modal information. To the best of our knowledge, this paper is the first to propose encoding motion patterns with varying temporal granularities into a shared discrete latent space as motion pattern tokens.
\item This work proposes Multi-Granular VQ-VAE (MGVQ-VAE), a novel model that can temporally tokenize motion patterns in multi-granularities using a shared codebook. Experiments validate the superiority of MGVQ-VAE compared with vanilla VQ-VAE.

\item Extensive experiments show that \textbf{M3G} outperforms state-of-the-art methods in generating audio-driven gestures, as confirmed by both objective evaluations and subjective human studies.

\end{itemize}

\section{Related Works}

\subsection{Human 3D Motion Generation Approaches}


Early attempts at generating human gestures involved using rule-based algorithms~\cite{cassell2001beat,huang2012robot} to select appropriate gestures from a database based on input and blend motion clips with smooth algorithms. However, due to limited data and varying speech details, these algorithms often resulted in out-of-sync speech and motion.


Deep generative models, such as Variational Autoencoders (VAEs)~\cite{kingma2013auto}, Generative Adversarial Networks (GANs)~\cite{goodfellow2014generative}, and transformers~\cite{vaswani2017attention,dehghani2018universal}, have shown promise in addressing the limitations of rule-based methods by capturing complex data correlations in a shared vector space. 
However, direct generation from a continuous latent space is sensitive to input noise, presenting challenges in practical applications. VAE-based methods also face the problem of "posterior collapse," restricting their ability to generate diverse gestures based on audio cues.


Recent work~\cite{yazdian2022gesture2vec,siyao2022bailando,ao2022rhythmic} has utilized VQ-VAE~\cite{van2017neural} to project motion patterns to a discrete latent codebook, transforming the motion generation problem into an auto-regressive token prediction task conditioned on audio. This approach enhances model robustness and generation quality. However, existing VQ-VAE methods assume each token represents poses of gestures in a single frame, limiting the generation of expressive motions varying in frame counts.


This paper introduces a variation of VQ-VAE called Multi-Granular VQ-VAE (MGVQ-VAE) to tokenize dynamic motions in different temporal granularities, enabling the generation of more detailed and natural gesture sequences.


\subsection{Audio-driven Full-body Gesture Generation}

Realistic and expressive gestures play a crucial role in enhancing virtual avatars' presence in the Metaverse. Inaccurate or illogical gestures can evoke the "Uncanny Valley" phenomenon, resulting in a negative user experience. Therefore, generating coherent and expressive full-body gestures is essential for maintaining user engagement.


Given the complexity and variability of human motions, recent research has focused on generating gestures for specific parts of the body, such as the face~\cite{fan2022faceformer,xing2023codetalker}, upper body (including arms, wrists, and hands)~\cite{yoon2020speech,ginosar2019learning}, and lower body (including overall movements)~\cite{yang2023diffusestylegesture} individually. While these generative models have shown promise in capturing unique motion patterns, they lack the versatility to generate movements across all body parts due to the diverse range of motion patterns among different body parts. There is a need for a unified framework for motion modeling.

A recent study~\cite{liu2023emage} approached generating full-body gestures by generating each part, the face, upper body, hands, lower body, and global movements first and combining these components to form full-body gestures. 

However, existing methods are limited by fixed granularity in modeling and generating motions, which restricts the ability to capture multi-granular motion patterns and extract semantic correlations between different body parts.

In this work, we utilize the multi-granular encoder network to extract the corase-to-fine-grained pattern information within the audio and transcript embeddings. It can predict the motion token series accurately by exploiting both the body part information and the semantic correlations between different body parts through a multi-granular temporal cross-attentive network.

\section{Methodology}

\begin{figure*}[htb]
\centering
\includegraphics[width=0.8\textwidth]{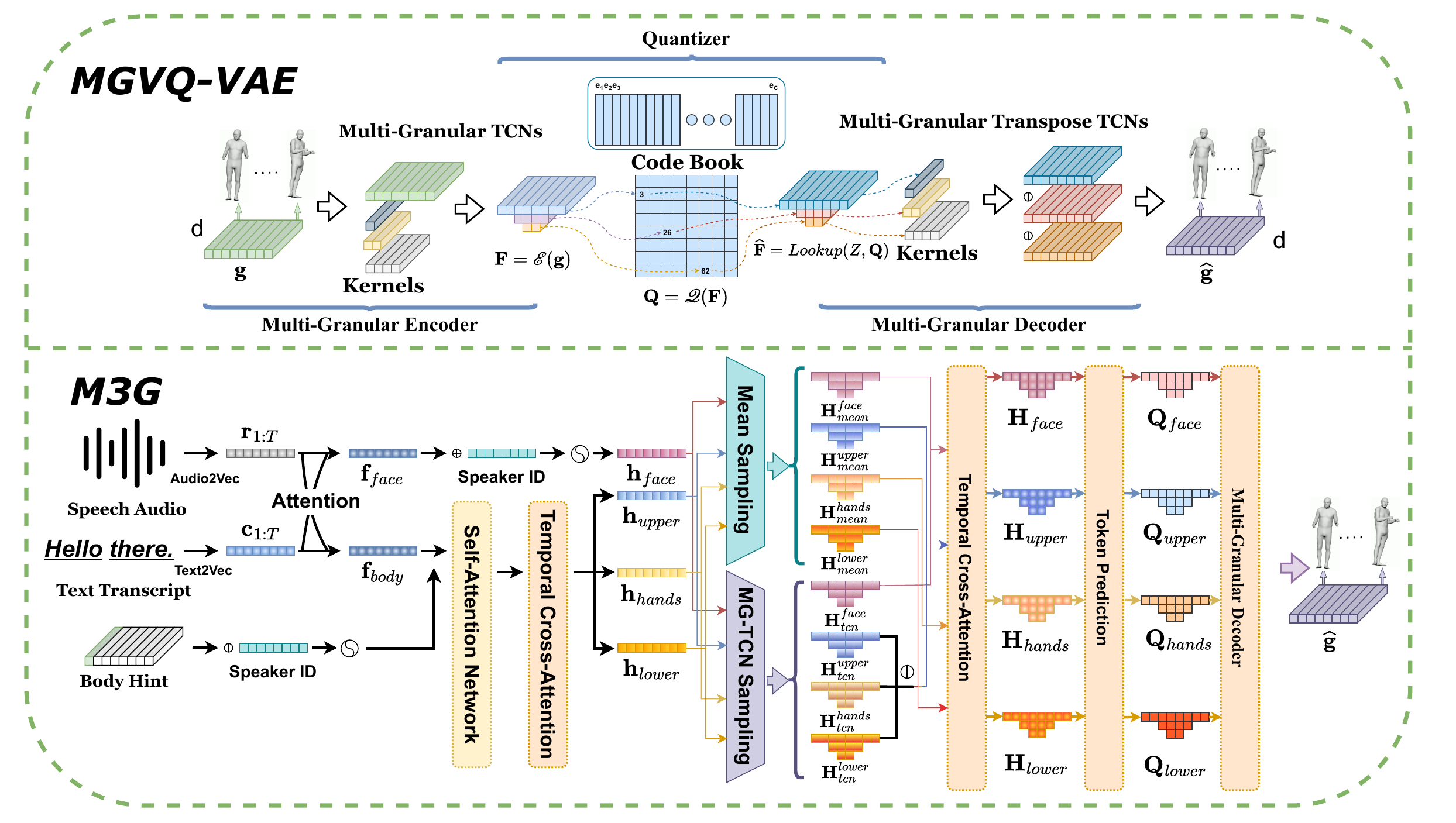} 
\caption{The Overall Workflow of M3G. We illustrate the workflow in two stages: 1. Training MGVQ-VAE to learn discrete latent representations of motions, as known as tokens, for encode and reconstruction. 2. Training Multi-Granular Token Predictor to predict the discrete latent representations from audio. }
\label{workflow}
\end{figure*}


This section presents the development of the Multi-Granular Gesture Generator (M3G) framework, as depicted in Figure~\ref{workflow}. Initially, we formally define the problem of audio-driven gesture generation. Subsequently, we outline the workflow of the M3G framework and delve into the structures of Multi-Granular VQ-VAE and Multi-Granular Token Predictor.

\subsection{Problem Formulation}
For a given audio $\mathbf{a}\in \mathbb{R}^{L\cdot sf}, sf=sr_{audio}/fps_{gestures}$, where $L$ represents the expected total frame counts of generated gesture sequences, $sf$ represents the sample count per frame, $sr_{audio}$ denotes the sample rate of the given audio, and $fps_{gestures}$ denotes the frames per second of the generate gesture sequences. 
The task is to generate the holistic 3D gesture of the human body  $\mathbf{g}\in \mathbb{R}^{L\times (55\times 6 + 100 + 4 + 3)}$ from the given audio. The dimension of $\mathbf{g}$ is formed by $L$ multiplied by the sum of 55 joints in Rot6D, 100 dimensions of facial expressions in FLAME parameters, 4 foot contact labels, and 3 global translation parameters.

\subsection{Multi-Granular Gesture Generator (M3G)}

The overall workflow of M3G can be divided into two stages: 1. training MGVQ-VAE to learn discrete latent representations of motions for encoding and reconstruction. 2. training Multi-Granular Token Predictor to predict the discrete latent representations from audio.



In the first stage, motion representations are tokenized at various granularities using multiple Temporal Convolution Networks (TCNs). These tokens are mapped to a shared codebook for semantic alignment. The original motions are reconstructed by applying Transpose TCN to each token series and aggregating the results.

In the second stage, the goal is to predict token series from input audio and text transcripts to reconstruct corresponding motion sequences. To account for the diverse motion patterns across different body parts, the audio and text representations are transformed into four separate latent embeddings for reconstructing the face, upper body, hands, and lower body. Multi-granular latent representations for each body part are generated using mean sampling and the MG-TCN sampling strategy. To capture correlations among the multi-granular latent representations of each body part, a temporal cross-attention mechanism is utilized.
Subsequently, all learned latent embeddings are quantized into token series of varying granularities and decoded by the multi-granular decoder in the trained MGVQ-VAE to reconstruct the motion sequences.


\subsection{Multi-Granular VQ-VAE (MGVQ-VAE)}

Normal VQ-VAE consists of Encoder $\mathcal{E}$, Quantizer $\mathcal{Q}$, and Decoder $\mathcal{D}$. The Encoder projects the raw gesture sequences ($\mathbf{g}$) into the latent representation space as encoded embeddings $\mathbf{f}$ frame by frame, and then the Quantizer classifies the embeddings into discrete tokens $\mathbf{q}$ according to the codebook $Z$ based on the embedding distance in the latent space. After that, the decoder reconstructs the motion sequences $\mathbf{\hat{g}}$ from the quantized embedding $\mathbf{\hat{f}}$, which is generated by looking up the codebook with quantized token series.
\begin{equation}
\begin{split}
    &\mathbf{f} = \mathcal{E}(\mathbf{g}), \mathbf{q} = \mathcal{Q}(\mathbf{f}), \mathbf{\hat{f}} = \mathrm{lookup}(Z, \mathbf{q}), \mathbf{\hat{g}} = \mathcal{D}(\mathbf{\hat{f}})\\
    &\mathbf{f}, \mathbf{\hat{f}} \in \mathbb{R}^{T_{source}\times d}, \mathbf{q} \in \mathbb{R}^{T_{source}}, Z \in \mathbb{R}^{C\times d}
\end{split}
\end{equation}

where $T_{source}$, $d$, and $C$ are the total frame count of the input motion sequence, the dimensions of embeddings, and the size of the codebook, respectively. 

In MGVQ-VAE, we utilize $n$ TCN networks representing $n$ levels of granularities to generate the encoded embedding $\mathbf{F}$, which consists of $n$ embedding series with various lengths. 
\begin{equation}
    \mathbf{F} = [\mathbf{f}_1, ..., \mathbf{f}_n] = \mathcal{E}(\mathbf{g}) = [TCN_1(\mathbf{g}), ..., TCN_n(\mathbf{g})]
\end{equation}

where $\mathbf{f}_i \in \mathbb{R}^{(T_{source}/{2^i})\times d}$ denotes the embedding series with $i$ from 1 to $n$.

In the quantization stage, the encoded embeddings are first quantized to token series $\mathbf{Q}$, then formed the quantized embedding series $\mathbf{\hat{F}}$ by looking up the codebook. All embedding series share the same codebook $Z$ to ensure consistency in the embedding semantics.
\begin{equation}
    \mathbf{Q} = [\mathbf{q}_1, ..., \mathbf{q}_n] = \mathcal{Q}(\mathbf{F}), \mathbf{\hat{F}} = \mathrm{lookup}(Z, \mathbf{Q})
\end{equation}

where $\mathbf{q}_i \in \mathbb{R}^{T_{source}/{2^i}}$ denotes the token series.

Specifically for the quantization function, we choose the token with the minimum $L_2$ distance between the encoded embedding and token-represented embedding on the codebook. The lookup function returns the token-represented embedding based on the input token according to the codebook. The element-wise functions are described in Eq. (4).
\begin{equation}
\begin{aligned}
    & q_k = \mathop{\arg\min}\limits_{j \in [0, C]} ||z_j - f_k||_2,\\
    &\hat{f}_k = z_k = \mathrm{lookup}(Z, q_k),\\
\end{aligned}
\end{equation}
where $z_j \in \mathbf{Z}$ denotes the token-represented embedding $j$ in codebook, $f_k$ denotes the $k^{th}$ encoded embedding in $i^{th}$ granularity in $\mathbf{F}$.

In the decoding stage, instead of using TCN as previous VQ-VAE-based methods~\cite{liu2023emage,ao2022rhythmic}, we utilize the Temporal Transpose Convolution Network (TransTCN) to decode the quantized embeddings into gesture sequences using the same scale of reception field in the encoder. Thus, multiple gesture sequences are reconstructed from quantized embeddings in different granularities. Then, we add them as the reconstructed gestures. 
\begin{equation}
\begin{aligned}
    \mathbf{\hat{g}} &= \mathcal{D}(\mathbf{\hat{F}}) \\
    &= \mathrm{sum}(\mathrm{TransTCN_1}(\mathbf{\hat{f}}_1), ..., \mathrm{TransTCN_n}(\mathbf{\hat{f}}_{n}))
\end{aligned}
\end{equation}



Notably, for each convolution operation in TCNs, the output embeddings are normalized by dividing the kernel size shown in Eq.(6) to mitigate the scale differences stemming from varying kernel sizes, thereby enhancing the stability of the training process.
\begin{equation}
    \mathbf{r} = Conv(\mathbf{f}, ks, s, p)/ks
\end{equation}

where $Conv()$ denotes convolutional operation, $ks$ denotes kernel size, $s$ denotes stride, and $p$ denotes padding.

The loss function of MGVQ-VAE is as follows:
\begin{equation}
\begin{aligned}
    &\mathcal{L}_{\mathrm{VQ-VAE}} = 
    \mathcal{L}_{rec}(\mathbf{g}, \mathbf{\hat{g}}) + \mathcal{L}_{vel}(\mathbf{g'}, \mathbf{\hat{g}'})  \\
    &+ \mathcal{L}_{acc}(\mathbf{g}'', \mathbf{\hat{g}''}) +||sg[\mathbf{F}]-\mathbf{\hat{F}}||_2^2 + ||\mathbf{F}-sg[\mathbf{\hat{F}}]||_2^2 \\
\end{aligned}
\end{equation}

where $ \mathcal{L}_{rec}$ is Geodesic~\cite{tykkala2011direct} loss, and $\mathcal{L}_{vel},  \mathcal{L}_{acc}$ are L1 losses. sg is the stop gradient operation and the weight of commitment~\cite{van2017neural} loss is set to 1.

\subsection{Multi-Granular Token Predictor}

In the Multi-Granular Token Predictor, we need to predict the corresponding multi-granular tokens series $\mathbf{\hat{Q}} = [\mathbf{\hat{q}}_1, ..., \mathbf{\hat{q}}_n]$ from audio features and text contents. 

\paragraph{Audio and Context Feature Fusion}
For the input audio features $\mathbf{s}$, following previous work~\cite{ao2023gesturediffuclip}, we employ onset $\mathbf{o}$, and amplitude $\mathbf{a}$ and combine them as rhythmic audio features $\mathbf{r} \in \mathbb{R}^{T\times d}$. For the input transcript text, we transform them into content features $\mathbf{c} \in \mathbb{R}^{T\times d}$ using pre-trained model~\cite{bojanowski2017enriching}.

Then we utilize the attention mechanism to fuse the audio features and context features 
\begin{equation}
\begin{aligned}
    & \alpha = \mathrm{Softmax}(MLP(\mathbf{r}_{1:T}, \mathbf{c}_{1:T})) \\
    & \mathbf{f}_{1:T} = \alpha \times \mathbf{r}_{1:T} + (1-\alpha)\times\mathbf{c}_{1:T} \\
\end{aligned}
\end{equation}

where $T$ denotes the total frame counts of the audio after sampling and the expected motion sequences. $\alpha \in \mathbb{R}^{T\times d}$ is the element-wise attention coefficient.

\paragraph{Latent Generation and Token Prediction}
According to the previous studies~\cite{liu2023emage}, in the holistic gesture generation task, there is only a weak correlation between the distribution of facial expression and body motion. Therefore, we propose to exploit two independent feature extraction and token prediction workflows for facial expression and body motion separately, based on independent audio-text representations $\mathbf{f_{face}}$ and $\mathbf{f_{body}}$.


For the latent representation prediction, we first combine the audio-text embedding with the learned speaker embedding and project to the dimension of the hidden vector:
\begin{equation}
\begin{aligned}
& \mathbf{h}_{face} = \mathrm{MLP}(\mathbf{f}_{face} \oplus \mathbf{p}_f) \\
& \mathbf{h}_{hints} = \mathcal{SAN}(\bar{\mathbf{g}} + \mathbf{p}_t)\\
& \mathbf{h}_{body} = \mathbf{h}_{hints} + \mathcal{TCAT}(\mathbf{h}_{hints}\oplus\mathbf{p}_t, \mathbf{f}_{body})\\
\end{aligned}
\end{equation}

$\mathbf{h}_{parts} \in \mathbb{R}^{T\times h}$ denotes the hidden representation of part motion, where $parts$ can be $face$ or $body$. $\mathbf{f}_{parts} \in \mathbb{R}^{T\times d}$ denotes the fused audio-text embedding, and $\mathbf{p}_f \in \mathbb{R}^{d}$ denotes the learned speaker embedding. Body hints $\mathbf{h}_{hints} \in \mathbb{R}^{T\times h}$ are encoded from the masked gesture sequence $\bar{g}$, where the first 8 frames are set to be the ground truth and the left frames are set to be 0 vectors. $\mathcal{SAN}$ denotes a self-attentive network.

To better capture the various motion patterns of different parts of the body, the hidden representation of the body is further projected into three different latent spaces, including the upper body, lower body, and hands. 
\begin{equation}
\begin{aligned}
& \mathbf{h}_{upper, lower, hands} = \mathrm{MLP}(\mathbf{h}_{body}) \\
\end{aligned}
\end{equation}

Then, we construct two types of multi-granular representations to generate multi-granular latent. The first type averages the consecutive representations as follows:
\begin{equation}
\begin{aligned}
& \mathbf{H}^{parts}_{mean} = [{\mathbf{h}^{parts}_{mean}}_0, ..., {\mathbf{h}^{parts}_{mean}}_{n-1}] \\
& {\mathbf{h}^{parts}_{mean}}_i = [\sigma(\mathbf{h}^{parts}_{[0,..,2^i-1]}),..., \sigma(\mathbf{h}^{parts}_{[n-2^i,..,n-1]})]\\
\end{aligned}
\end{equation}

where $parts$ can be $face$, $upper$, $lower$, and $hands$. $\sigma$ denotes the average operation and ${\mathbf{h}^{parts}_{mean}}_i \in \mathbb{R}^{T/2^{i} \times h}$ denotes the $i^{th}$ embedding series with $2^i$ frames of granularity.

The second type is the isomorphic TCN network in the MGVQ-VAE encoder with different input dimensions. This network can extract similar information as the MGVQ-VAE encoder and generate multi-granular motion latent.
\begin{equation}
\begin{aligned}
\mathbf{H}^{parts}_{tcn} 
& = [\mathcal{TCN}_0(\mathbf{h}_{parts}),...,\mathcal{TCN}_{n-1}(\mathbf{h}_{parts})] \\
& = [{\mathbf{h}^{parts}_{tcn}}_0, ..., {\mathbf{h}^{parts}_{tcn}}_{n-1}]\\
\end{aligned}
\end{equation}
where ${\mathbf{h}^{parts}_{tcn}}_i \in \mathbb{R}^{T/2^{i} \times h}$ denotes the $i^{th}$ embedding series with $2^i$ frames of granularity.

After constructing the embedding series with different granularities, a temporal cross-attention Transformer decoder $\mathcal{TCAT}$ is applied to capture the correlations between the above two representations. Then a 1-layer MLP is used to project the dimensions into the reconstruction latent $\mathbf{\hat{H}}^{parts}_{rec}$.
\begin{equation}
\begin{aligned}
& \mathbf{\hat{H}}^{parts}_{rec} = \mathrm{MLP}(\mathcal{TCAT}(\mathbf{H}^{parts}_{mean}, \mathbf{H}^{parts}_{tcn}))\\
\end{aligned}
\end{equation}

We optimize the learned reconstruction latent by applying the MSE loss:
\begin{equation}
\begin{aligned}
& \mathcal{L}^{parts}_{rec} = MSE Loss(\mathbf{\hat{H}}^{parts}_{rec}, \mathbf{\hat{F}}^{parts} )\\
\end{aligned}
\end{equation}

where $\mathbf{\hat{F}}^{parts}$ is the learned latent representation of MGVQ-VAE of corresponding body parts from ground truth body motion.

After learning the mutual information between two different types of representations, we further learned the multi-granular correlation among body parts. We added the TCN latent series of all body parts together to form a full-body multi-granular latent series and conducted cross-attention operations with the mean latent series: 
\begin{equation}
\begin{aligned}
\mathbf{H}_{full} = & \mathbf{H}^{upper}_{tcn} + \mathbf{H}^{hands}_{tcn} + \mathbf{H}^{lower}_{tcn}\\
\mathbf{\tilde{H}}_{parts} 
& = \mathcal{TCAT}(\mathbf{H}^{parts}_{mean}, \mathbf{H}_{full})\\
& = [\mathbf{\tilde{h}}^{parts}_0, ..., \mathbf{\tilde{h}}^{parts}_{n-1}]\\
\end{aligned}
\end{equation}
where $\mathbf{\tilde{H}}_{parts}$ denotes the latent embedding of each part of the body in different granularities and $parts$ can be $upper$, $lower$, and $hands$ according to previous body part separation. $\mathbf{h}^{parts}_i \in \mathbb{R}^{T/2^i \times h}$ represents the latent embedding of body parts in different granularities.

After generating the latent representations of each body part, we quantize the latent representations by predicting the corresponding tokens iteratively from coarse to fine granularity. Such quantization operation projects the learned continuous latent to discrete latent space represented by tokens for better reconstruction performance.
\begin{equation}
\begin{aligned}
& \mathbf{\hat{q}}^{parts}_i = \mathrm{MLP}(\mathbf{\tilde{h}}^{parts}_i + \mathbf{\hat{q}}^{parts}_{i+1}), i \in [0, n-2]\\
& \mathbf{\hat{Q}}_{parts} = [\mathbf{\hat{q}}^{parts}_0,...,\mathbf{\hat{q}}^{parts}_{n-1}]\\
\end{aligned}
\end{equation}
where $\mathbf{\hat{q}}^{parts}_i \in \mathbb{R}^{T/2^i \times C}$ represents the predicted possibility distribution of motion tokens for $i^{th}$ granularity, $C$ denotes the codebook size.

We optimize the learned token index distribution by applying the cross-entropy loss:
\begin{equation}
\begin{aligned}
& \mathcal{L}^{parts}_{cls} = \mathrm{CrossEntropy}(\mathbf{\hat{Q}}_{parts}, \mathbf{Q}_{parts} )\\
\end{aligned}
\end{equation}

where $\mathbf{Q}_{parts}$ is the learned token of body motion in MGVQ-VAE.

\section{Experiments}
\subsection{Datasets}
We evaluate the ability of our method to generate holistic 3D gestures from speech on a diverse and expressive dataset BEAT2\footnote{https://github.com/PantoMatrix/PantoMatrix}~\cite{liu2022beat} collected from mocap equipment. This public dataset contains 76 hours of high-quality, multi-modal data captured from 30 speakers talking with eight different emotions.
Following the settings of existing work~\cite{liu2023emage}, we conduct the experiments on the BEAT2-Standard Speaker2 with an 85\%/7.5\%/7.5\% train/val/test split.

\subsection{Evaluation Metrics}

For the body motion generation, we adopt \textbf{FGD}~\cite{yoon2020speech} to measure the similarity between the generated gesture and the real gestures. To evaluate the \textbf{Diversity}~\cite{li2021audio2gestures} of generated gestures, we calculate the L1 distance between different gesture clips. In terms of audio-motion synchronization, we utilize the \textbf{Beat Align}~\cite{li2021ai} measurement. 

For the facial expression generation, we use \textbf{Mean Square Error (MSE)}~\cite{xing2023codetalker} to measure the vertex position distance and 
\textbf{L1 Vertex Difference (LVD)}~\cite{xing2023codetalker} to measure the L1 distance between the generated facial expression and the ground truth facial expression.

The results are reported as the mean value and the standard deviation computed from 5 times of independent runs. The significance is also reported with the p-value.

\subsection{Comparison Methods}

We compare our M3G\footnote{The source code will be released to GitHub after acceptance.} with the following classic and state-of-the-art methods of talking head generation and body gesture generation: 
S2G~\cite{ginosar2019learning}, Trimodal~\cite{yoon2020speech}, HA2G~\cite{liu2022learning}, DisCo~\cite{liu2022disco}, CaMN~\cite{liu2022beat}, Diffusestylegesture~\cite{yang2023diffusestylegesture}, Habibie et al~\cite{habibie2021learning},TalkSHOW~\cite{yi2023generating}, and EMAGE~\cite{liu2023emage}.

For Habibie et al~\cite{habibie2021learning}, TalkSHOW~\cite{yi2023generating}, we used their original version, which is the upper body generation. Meanwhile, we cite the results of their full-body version reproduced by ~\cite{liu2023emage}.

\subsection{Overall Comparison}
In this part, we will compare the overall performances of M3G with classical and state-of-the-art audio-driven gesture generation methods. In Table~\ref{tab:Overall Performance}, \textbf{Habibie et al$^*$} and \textbf{TalkSHOW$^*$} denotes the reported performance of reproduced full-body motion generation in ~\cite{liu2023emage}, thus no std values are reported.

\begin{table}[htb!]
\centering
\resizebox{\columnwidth}{!}{
\setlength{\tabcolsep}{1.0mm}{
\begin{tabular}{l|ccccc}
    & 
    {\textbf{FGD $^{\times 10^{-1}}$$\downarrow$}} &
    {\textbf{BA$^{\times 10^{-1}}$$\uparrow$}} &{\textbf{Diversity$\uparrow$}} &{\textbf{MSE$^{\times 10^{-8}}$$\downarrow$}} & {\textbf{LVD$^{\times 10^{-5}}$$\downarrow$}}\\
    \midrule
     \textbf{S2G} & 27.87 $\pm$ 0.343 & 4.827 $\pm$ 0.138 & 6.022 $\pm$ 0.097 & - & - \\
     
     \textbf{Trimodal} & 12.13 $\pm$ 0.141 & 5.762 $\pm$ 0.063 & 7.513 $\pm$ 0.073 & - & -  \\
     
     \textbf{HA2G} & 12.32 $\pm$ 0.053 & 6.779 $\pm$ 0.021 & 8.626 $\pm$ 0.016& - & -  \\
    
     \textbf{DisCo} & 9.484 $\pm$ 0.066 & 6.439 $\pm$ 0.027 & 9.912 $\pm$ 0.022 & - & -  \\
    
     \textbf{CaMN} & 6.967 $\pm$ 0.023 & 6.628 $\pm$ 0.018 & 11.18 $\pm$ 0.089 & - & -  \\
    
     \textbf{DiffStyleGesture} & 8.866 $\pm$ 0.243 & 7.239 $\pm$ 0.089 & 11.13 $\pm$ 0.077 & - & -  \\

     \textbf{Habibie et al} & 8.869 $\pm$ 0.014 & \textbf{7.810 $\pm$ 0.005} & 8.372 $\pm$ 0.065 & 8.614 $\pm$ 0.026 & 8.043 $\pm$ 0.033 \\
     
     \textbf{Habibie et al$^*$} & 9.040 & 7.716 & 8.213 & 8.614 & 8.043  \\

     \textbf{TalkSHOW} & 6.145 $\pm$ 0.011 & 6.863 $\pm$ 0.008& 13.12 $\pm$ 0.156 & 7.791 $\pm$ 0.044 & 7.771 $\pm$ 0.052 \\
     
     \textbf{TalkSHOW$^*$} & 6.209 & 6.947 & 13.47 & 7.791 & 7.771 \\

     \textbf{EMAGE} & 5.643 $\pm$ 0.015 & 7.707 $\pm$ 0.004& 12.92 $\pm$ 0.198 & 7.694 $\pm$ 0.076 & 7.593 $\pm$ 0.062  \\
     
      \textbf{M3G} & \textbf{4.784 $\pm$ 0.014} & 7.398 $\pm$ 0.008 & \textbf{13.53 $\pm$ 0.020} & \textbf{7.291 $\pm$ 0.028} & \textbf{7.439 $\pm$ 0.011}  \\
\midrule
    \textit{p-value} & $< 0.0001$ & $< 0.0005$ & $< 0.0001$ & $< 0.0001$ & $< 0.001$ 
\end{tabular}}}
\caption{Overall Comparison of various methods.  The best performance of each metric is in \textbf{boldface} fonts. The sign $\uparrow$ beside the metric denotes that the larger the value, the better it is, while the sign $\downarrow$ is the reverse.}
\label{tab:Overall Performance}
\end{table}
Table~\ref{tab:Overall Performance} reported that our proposed \textbf{M3G} outperforms the state-of-the-art methods on nearly all evaluation metrics. Especially, \textbf{M3G} significantly improves the reconstruction accuracy performances such as \textbf{FGD}, \textbf{MSE}, and \textbf{LVD} compared with the state-of-the-art method, demonstrating that the generated full-body gestures are closer to the ground-truth motion. Such results indicate that by extending the latent representation to multi-granularity, \textbf{M3G} is capable of modeling a greater variety of motion patterns, which facilitates the model's ability to generate motion sequences with better quality and accuracy than other methods.

\subsection{Ablation Analysis}
\subsubsection{Granularity Number Selection}
In this section, we will evaluate the impact of the level of granularities on the motion reconstruction performance. 

The granularity level increases following the size list [1, 2, 4, 8, 16] to prevent information overlap among different sizes. In this experiment, we adjust the number of granularities used to find the best one. For instance, setting 4 granularities represents using 4 increasing granularity sizes from size 1 to form [1, 2, 4, 8] as the M3G model's granularity level. The experiment results are shown in Table~\ref{tab:Granularity Selection}

\begin{table}[htb!]
\centering
\resizebox{\columnwidth}{!}{
\setlength{\tabcolsep}{1.0mm}{
\begin{tabular}{c|ccccc}

    & 
    {\textbf{Face$^{\times 10^{-3}}$$\downarrow$}} &
    {\textbf{Upper$^{\times 10^{-2}}$$\downarrow$}} &{\textbf{Hands$^{\times 10^{-2}}$$\downarrow$}} &{\textbf{Lower$^{\times 10^{-2}}$$\downarrow$}} & {\textbf{Global$^{\times 10^{-2}}$$\downarrow$}}\\
    \midrule
     
      \textbf{1 Granularity} & 1.883 $\pm$ 0.033 & 3.482 $\pm$ 0.016 & 6.843 $\pm$ 0.006 & 2.648 $\pm$ 0.017 & 4.752 $\pm$ 0.016  \\
     
      \textbf{2 Granularities} & 1.667 $\pm$ 0.026 & 3.262 $\pm$ 0.022 & 5.949 $\pm$ 0.015 & 2.546 $\pm$ 0.027 & 4.783 $\pm$ 0.009  \\
     
      \textbf{3 Granularities} & 1.552 $\pm$ 0.038  & 2.993$\pm$ 0.025  & 5.218 $\pm$ 0.021 & 2.484 $\pm$ 0.035 & 4.421 $\pm$ 0.025  \\

      \textbf{4 Granularities} & \textbf{1.368 $\pm$ 0.031}  & 2.972$\pm$ 0.020  & \textbf{4.818 $\pm$ 0.024} & \textbf{2.302 $\pm$ 0.038} & \textbf{4.103 $\pm$ 0.017}  \\

      \textbf{5 Granularities} & 1.508 $\pm$ 0.022  & \textbf{2.612$\pm$ 0.023}  & 5.483 $\pm$ 0.014 & 2.339 $\pm$ 0.029 & 4.611 $\pm$ 0.023  \\

          \midrule

    \textit{p-value} & $< 0.0001$ & $< 0.0001 $ & $< 0.0005$ & $< 0.0001$ & $< 0.0001$ \\
    \midrule
    \midrule
    & 
    {\textbf{FGD$^{\times 10^{-1}}$$\downarrow$}} &
    {\textbf{BC$^{\times 10^{-1}}$$\uparrow$}} &{\textbf{Diversity$\uparrow$}} &{\textbf{MSE$^{\times 10^{-8}}$$\downarrow$}} & {\textbf{LVD$^{\times 10^{-5}}$$\downarrow$}}\\
    \midrule
        \textbf{1 Granularity} & 8.821 $\pm$ 0.028 & 7.122 $\pm$ 0.014 & 13.352 $\pm$ 0.007 & 10.363 $\pm$ 0.004 & 8.728 $\pm$ 0.008  \\
     
        \textbf{2 Granularities} & 6.768 $\pm$ 0.036 & 7.342 $\pm$ 0.027 & 12.979 $\pm$ 0.005 & 8.211 $\pm$ 0.005 & 7.858 $\pm$ 0.010  \\    
     
        \textbf{3 Granularities} & 5.627 $\pm$ 0.017 & 7.254 $\pm$ 0.012 & \textbf{13.723 $\pm$ 0.011} & 8.091 $\pm$ 0.010 & 7.816 $\pm$ 0.016  \\

        \textbf{4 Granularities} & \textbf{4.784 $\pm$ 0.014} & \textbf{7.398 $\pm$ 0.008} & 13.530 $\pm$ 0.020 & \textbf{7.291 $\pm$ 0.028} & \textbf{7.439 $\pm$ 0.011} \\

        \textbf{5 Granularities} & 6.463 $\pm$ 0.017 & 6.982 $\pm$ 0.012 & 12.263 $\pm$ 0.008 & 8.889 $\pm$ 0.005 & 8.258 $\pm$ 0.012  \\
    \midrule

    \textit{p-value} & $< 0.0001$ & $< 0.0001 $ & $< 0.0001$ & $< 0.0001$ & $< 0.0001$ \\
\end{tabular}}}
\caption{Experiments for Granularity Selection}
\label{tab:Granularity Selection}
\end{table}

The experiment results show that increasing the granularity number initially boosts performance but eventually leads to a decline. A higher number allows M3G to better capture continuous motion patterns by using hierarchical granularity tokens, enhancing gesture fluency and realism. However, a larger granularity size does not continually lead to better performance. It might be caused by the overcomplexity of longer clips, which increases tokenize complexity. Moreover, adding more granularity can even negatively impact the results due to the M3G’s limited token prediction ability.

\subsubsection{Key components}
In this section, we will evaluate the contributions of the multi-granularity mechanism and each component proposed in this work to the superiority of \textbf{M3G}. 

Thus, we propose five variants of \textbf{M3G}: \textbf{w/o Separate Body Latents}, which use unified latent embeddings for different body parts instead of separate latent representations. \textbf{w/o Text}, which only uses audio information to generate motion sequences. \textbf{w/o TCN Encoder}, which disabled the TCN network in equation (12) that extracts multi-granular features in the token predictor, \textbf{w/o Full-Body Latent}, which disabled the full-body multi-granular latent series involvement in equation (15) and generated each body parts' latent embedding only based on their own parts' features, \textbf{w/o TransTCN}, which substituted the TransTCN structure in equation (4) with the TCN structure in equation (2) following previous work~\cite{liu2023emage,ao2022rhythmic}.

Table~\ref{tab:Ablation Experiments} reported that the \textbf{M3G} significantly outperforms or performs similarly with all variants, demonstrating the contributions of these components. Moreover, the results indicate that replacing the \textbf{TransTCN} structure leads to a more significant decline in performance compared with the other two variants, demonstrating the indispensable of the token decoding process in the overall workflow. 
The absence of \textbf{Full-Body latent} mainly affects the body's reconstructing performance, which might be caused by the lack of mutual information among different body parts.
The \textbf{w/o TCN Encoder} performs significantly worse than \textbf{M3G} on facial reconstruction and body diversity, indicating the ability of multi-granular TCN Encoder to model diverse types of motion patterns and facial expressions from audio.

\begin{table}[htb!]
\centering
\resizebox{\columnwidth}{!}{
\setlength{\tabcolsep}{1.0mm}{
\begin{tabular}{l|ccccc}

    & 
    {\textbf{FGD$^{\times 10^{-1}}$$\downarrow$}} &
    {\textbf{BC$^{\times 10^{-1}}$$\uparrow$}} &{\textbf{Diversity$\uparrow$}} &{\textbf{MSE$^{\times 10^{-8}}$$\downarrow$}} & {\textbf{LVD$^{\times 10^{-5}}$$\downarrow$}}\\
    \midrule

      \textbf{w/o Separate Body Latents} & 8.926 $\pm$ 0.021 & 6.843 $\pm$ 0.008 & 10.838 $\pm$ 0.024 & 9.243 $\pm$ 0.013 & 8.153 $\pm$ 0.019 \\

      \textbf{w/o Text} & 6.677 $\pm$ 0.023 & 7.374 $\pm$ 0.012 & 12.893 $\pm$ 0.043 & 8.612 $\pm$ 0.012 & 7.852 $\pm$ 0.011 \\

      \textbf{w/o TransTCN} & 6.178 $\pm$ 0.035 & 7.132 $\pm$ 0.005 & 12.869 $\pm$ 0.057 & 8.833 $\pm$ 0.017 & 9.511 $\pm$ 0.015 \\
     
      \textbf{w/o Full-Body Latent} & 5.152 $\pm$ 0.034 & 7.232 $\pm$ 0.002 & 13.308 $\pm$ 0.062 & 7.440 $\pm$ 0.013 & 7.518 $\pm$ 0.012 \\
     
      \textbf{w/o TCN Encoder} & 5.178 $\pm$ 0.013 & \textbf{7.436} $\pm$ 0.004 & 12.439 $\pm$ 0.117 & 7.607 $\pm$ 0.081 & 7.589 $\pm$ 0.039  \\
     
      \textbf{M3G} & \textbf{4.784 $\pm$ 0.014} & 7.398 $\pm$ 0.008 & \textbf{13.53 $\pm$ 0.020} & \textbf{7.291 $\pm$ 0.028} & \textbf{7.439 $\pm$ 0.011}  \\
      \midrule

    \textit{p-value} & $< 0.0005$ & $> 0.1 $ & $< 0.0001$ & $< 0.0001$ & $< 0.0001$ \\
      
\end{tabular}}}
\caption{Ablation Experiments for Proposed Components.}

\label{tab:Ablation Experiments}
\end{table}
\subsection{MGVQ-VAE Reconstruction Performance}

This section evaluates the performance of \textbf{MG-VQVAE} in learning and reconstructing compared to \textbf{VQ-VAE} in existing methods. A variant of MG-VQVAE called \textbf{Unscaled MG-VQVAE} is also introduced, which excludes the scaling operation in Eq.(7) to examine its necessity.


Table~\ref{tab:Reconstruction Performance} presents the joints' rotation Mean Square Error (JRMSE) for each body part compared to the ground truth sequences. The second part of the table shows the metrics used in Table~\ref{tab:Overall Performance} to assess the reconstruction performance based on the encoded tokens in MG-VQVAE.
\begin{table}[htb!]
\centering
\resizebox{\columnwidth}{!}{
\setlength{\tabcolsep}{1.0mm}{
\begin{tabular}{l|ccccc}
    & 
    {\textbf{Face$^{\times 10^{-3}}$$\downarrow$}} &
    {\textbf{Upper$^{\times 10^{-2}}$$\downarrow$}} &{\textbf{Hands$^{\times 10^{-2}}$$\downarrow$}} &{\textbf{Lower$^{\times 10^{-2}}$$\downarrow$}} & {\textbf{Global$^{\times 10^{-2}}$$\downarrow$}}\\
    \midrule
     
      \textbf{VQ-VAE} & 2.100 $\pm$ 0.009 & 5.209 $\pm$ 0.028 & 7.103 $\pm$ 0.017 & 3.335 $\pm$ 0.023 & 4.418 $\pm$ 0.011  \\
     
      \textbf{Unscaled MGVQ-VAE} & 1.912 $\pm$ 0.075 & 3.181 $\pm$ 0.030 & 4.989 $\pm$ 0.055 & 2.991 $\pm$ 0.034 & 4.331 $\pm$ 0.047  \\
     
      \textbf{MGVQ-VAE} & \textbf{1.368 $\pm$ 0.022}  & \textbf{2.972$\pm$ 0.016}  & \textbf{4.818 $\pm$ 0.024} & \textbf{2.302 $\pm$ 0.025} & \textbf{4.103 $\pm$ 0.020}  \\

          \midrule

    \textit{p-value} & $< 0.0001$ & $< 0.0001 $ & $< 0.0005$ & $< 0.0001$ & $< 0.0001$ \\
    \midrule
    \midrule
    & 
    {\textbf{FGD$^{\times 10^{-1}}$$\downarrow$}} &
    {\textbf{BC$^{\times 10^{-1}}$$\uparrow$}} &{\textbf{Diversity$\uparrow$}} &{\textbf{MSE$^{\times 10^{-8}}$$\downarrow$}} & {\textbf{LVD$^{\times 10^{-5}}$$\downarrow$}}\\
    \midrule
      \textbf{VQ-VAE} & 3.302 $\pm$ 0.036 & \textbf{7.488 $\pm$ 0.014} & 12.482 $\pm$ 0.009 & 0.524 $\pm$ 0.002 & 2.087 $\pm$ 0.010  \\
     
      \textbf{Unscaled MGVQ-VAE} & 3.029 $\pm$ 0.036 & 6.949 $\pm$ 0.027 & 12.549 $\pm$ 0.005 & 0.531 $\pm$ 0.003 & 2.088 $\pm$ 0.006  \\    
     
      \textbf{MGVQ-VAE} & \textbf{1.497 $\pm$ 0.017} & 7.143 $\pm$ 0.012 & \textbf{12.654 $\pm$ 0.008} & \textbf{0.441 $\pm$ 0.005} & \textbf{2.011 $\pm$ 0.012}  \\
    \midrule
    \textit{p-value} & $< 0.0001$ & $< 0.0001 $ & $< 0.0001$ & $< 0.0001$ & $< 0.0001$ \\
\end{tabular}}}
\caption{Experiments for Reconstruction Errors.}
\label{tab:Reconstruction Performance}
\end{table}
The results show that our proposed MGVQ-VAE surpasses the traditional VQ-VAE in motion pattern tokenization and reconstruction. MGVQ-VAE effectively encodes motion patterns into multiple granular tokens, preserving more information than existing VQ-VAE. 

Furthermore, MGVQ-VAE outperforms the Unscaled version of MGVQ-VAE, highlighting the importance of the scaling operation in regularizing the scale of the predicted latent vector influenced by varying kernel sizes. This reflects that the regularization ensures all token embeddings with different granularities are distributed in the same semantic space represented by the MGVQ-VAE codebook.

\section{Perceptual Study}

To evaluate the naturalness of our generated results from the subjective perception, we conducted A/B testing following~\cite{yi2023generating} by comparing the generated motions by M3G with other methods including \textbf{CaMN}, \textbf{EMAGE}, \textbf{MGVQ-VAE Reconstruction}, and \textbf{Ground Truth}. For \textbf{CaMN}, since it does not generate the facial expression, we directly use the ground truth as its facial expression. 

We randomly sampled 40 clips from the generated video and compared them pairwisely. Eighteen participants took part in this study. Specifically, participants are asked to answer A or B to the following questions: Which clip do you think is more natural? 
\begin{figure}[htb]
\centering
\includegraphics[width=0.6\linewidth]{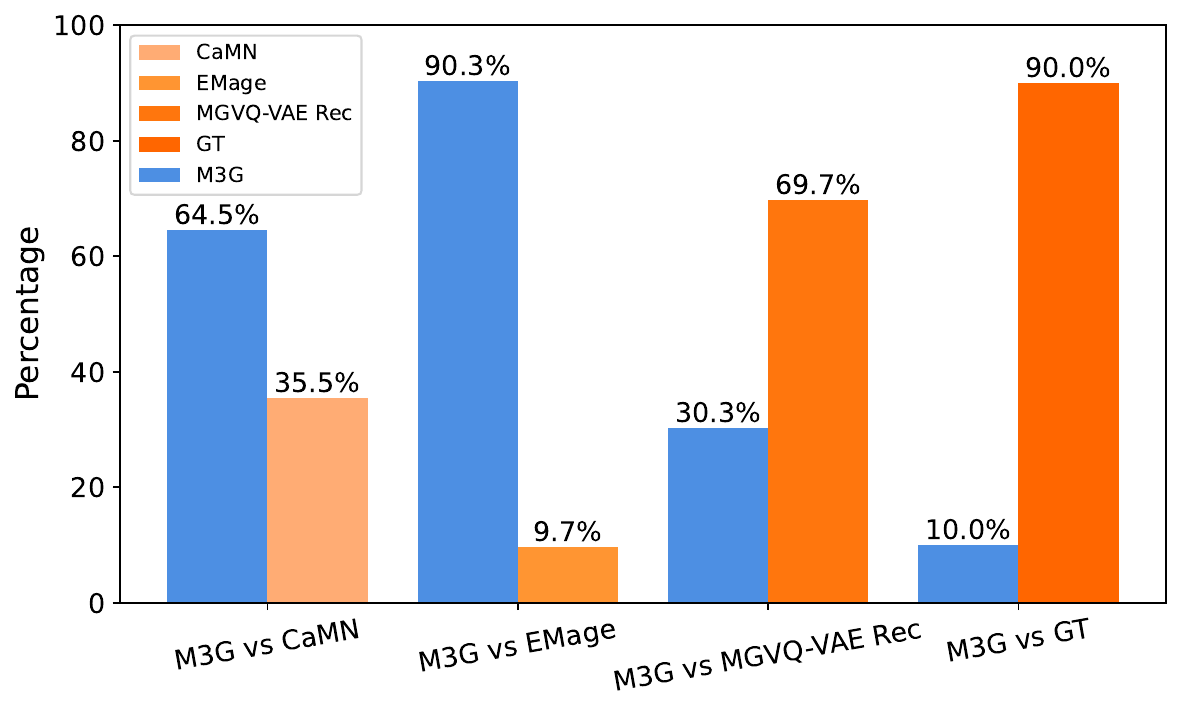} 
\caption{Perceptual study results on motion generation. }
\label{perceptual study}
\end{figure}
\begin{figure}[htb]
\centering
\includegraphics[width=\linewidth]{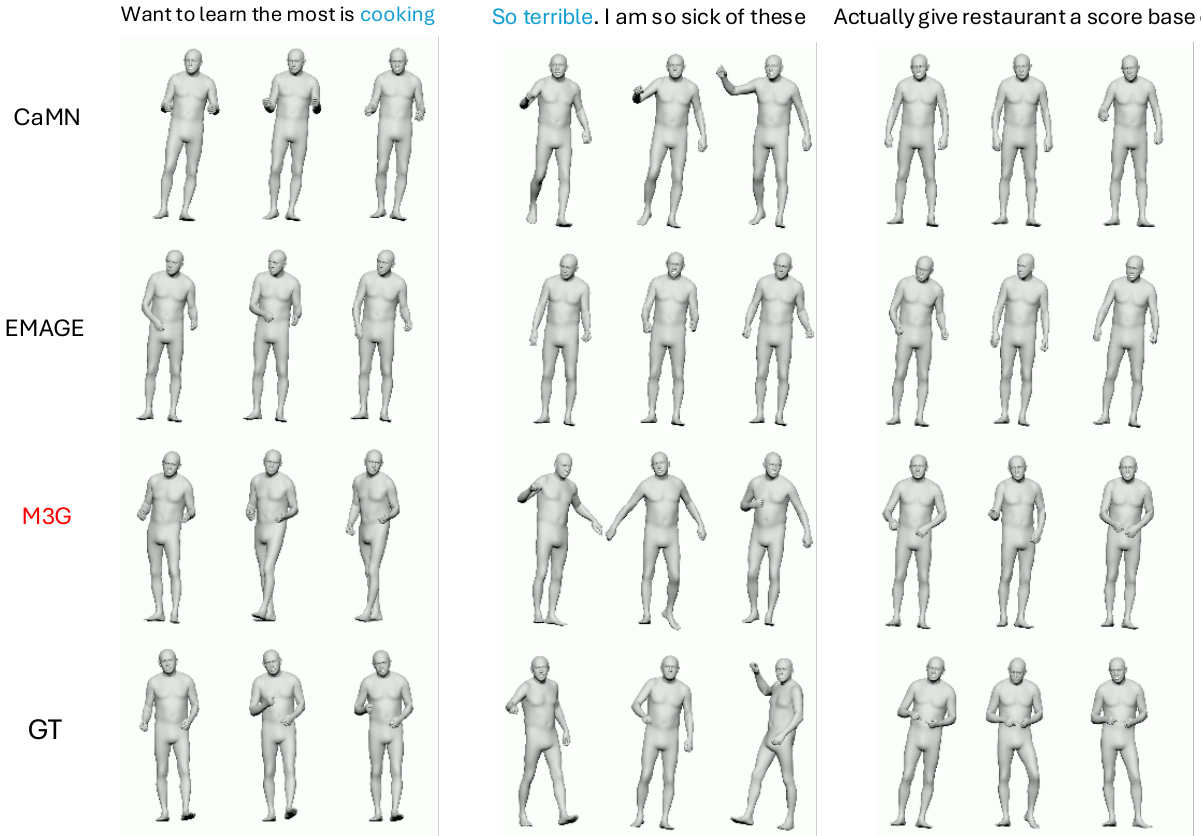} 
\caption{Qualitative results. }
\label{Qualitative results}
\end{figure}
Figure~\ref{perceptual study} illustrates that participants favor \textbf{M3G} over existing work \textbf{CaMN}, and \textbf{EMAGE}, indicating the superiority of \textbf{M3G} in generating more natural motion from audio than existing work. Moreover, in a proportion of cases, \textbf{M3G} out-performs the \textbf{MGVQ-VAE Reconstruction}, which generates the motion from encoded tokens in MGVQ-VAE, showing the closeness of generation quality between them. And \textbf{Ground Turth} outperforms \textbf{M3G} without doubt showcasing the potential for future developments. 

\section{Conclusion}
This study introduces an audio-driven holistic gesture synthesis framework named \textbf{M3G}, which consists of two models, MGVQ-VAE and multi-granular token predictor, outperforms other exiting methods in terms of facial expression generation, body motion generation and diversity of generated gestures. 
Our findings indicate that the proposed MGVQ-VAE effectively captures and preserves motion pattern information through multi-granular tokens, as evidenced by the superior reconstruction performance compared with vanilla VQ-VAE. By cooperating MGVQ-VAE with the multi-granular token predictor, which extracts multi-granular information from audio and predicts corresponding tokens for full-body motion reconstruction, \textbf{M3G} generates audio-driven gestures that accurately express semantic information. Extensive experiments demonstrate that our framework, M3G, generates more natural and realistic holistic 3D gestures synchronized with speech input, outperforming current state-of-the-art methods.




\appendix


\end{document}